# Machine-learning interatomic potential for AlN for epitaxial simulation

Nicholas Taormina[1*], Emir Bilgili[1], Jason Gibson[2], Richard Hennig[2], Simon Phillpot[2], Youping Chen[1]

[1]Department of Mechanical and Aerospace Engineering, University of Florida, Gainesville, Florida, 32611 USA

[2]Department of Materials Science and Engineering, University of Florida, Gainesville, Florida, 32611 USA

## ABSTRACT

A machine learned interatomic potential for AlN was developed using the ultra-fast force field (UF3) methodology. A strong agreement with density functional theory calculations in predicting key structural and mechanical properties, including lattice constants, elastic constants, cohesive energy, and surface energies has been demonstrated. The potential was also shown to accurately reproduce the experimentally observed atomic core structure of edge dislocations. Most significantly, it reproduced the experimentally observed wurtzite crystal structure in the overlayer during homoepitaxial growth of AlN on wurtzite AlN, something that prior potentials failed to achieve. Additionally, the potential reproduced the experimentally observed layer-by-layer growth mode in the epilayer. The combination of accuracy, transferability, and computational speed afforded by the UF3 framework thus makes large-scale, atomistic simulations of epitaxial growth of AlN feasible.

* Corresponding Author: Nicholas A. Taormina, ntaormina@ufl.edu

# I. Introduction

Aluminum Nitride (AlN) is a wide bandgap semiconductor that has gained significant attention for its combination of desirable thermal, mechanical, and electrical properties [1, 2]. These properties make AlN a key material in advanced electronic applications, particularly when used in conjunction with other semiconductors. While gallium nitride (GaN) remains the dominant active material for next-generation transistors [3], AlN is indispensable to GaN-based heterostructures. The close match in lattice parameter and thermal expansion coefficient with GaN make it the standard buffer layer for heteroepitaxial growth, where it mitigates strain, suppresses cracking, and reduces dislocation density in the overlayer [4-7]. Furthermore, strong piezoelectric polarization at the AlN/GaN interface gives rise to a highly conductive two-dimensional electron gas, enabling exceptional transport properties observed in GaN-based high-electron-mobility transistors (HEMTs) [8, 9].

However, the introduction of an AlN buffer layer brings its own set of challenges. Dislocation formation within the AlN layer can propagate into the GaN overlayer, degrading crystalline quality, scattering electrons and phonons, and ultimately limiting device performance and reliability [10, 11]. Since epitaxy is currently the only tool for manufacturing nitrides such as GaN and AlN and their heterostructures [12], understanding the mechanisms that drive the formation and evolution of dislocations during epitaxy is therefore critical. Dislocations arise from strain relaxation mechanisms such as dislocation nucleation or misorientation during island coalescence [13, 14]. These processes are inherently highly nonequilibrium and multiscale, coupling surface kinetics, adatom diffusion, and elastic strain fields. Because dislocation networks and extended defects emerge spontaneously from the growth processes, they cannot be manually introduced into the model but rather must be obtained through the simulation of epitaxial growth. The large length scales involved in the epitaxial processes make it intractable for first-principles methods. Atomic-level molecular dynamics or coarse-grained atomistic simulations provide a promising path to study the mechanisms behind dislocation formation in epitaxy. However, epitaxy is a highly nonequilibrium process and requires the interatomic potentials to be highly accurate in reproducing the atomic interactions, and hence the processes of crystallization and defect formation.

Many interatomic potentials for the AlN or GaN system have been developed over the years, with the goals to reproduce the structural, mechanical, or thermal transport properties of the system [15-19]. Despite extensive research efforts, the capability of existing empirical interatomic potentials in predictive simulation of nitride epitaxy remains limited. Specifically, to our knowledge, none of empirical potentials have been successful in reproducing the wurtzite structure observed in the grown overlayer of AlN or GaN epitaxy [20].

The structural and energetic characteristics of surfaces and interfaces differ fundamentally from those of bulk systems. These regions are marked by atomic-scale irregularities, broken symmetry, and complex local bonding environments. Interfaces often introduce strain or defects due to lattice mismatch. Capturing these phenomena requires a modeling approach that can accurately represent complex potential energy surfaces. Traditional empirical force fields generally fall short in this regard due to their unflexible functional forms, which limits transferability. Machine-learned interatomic potentials (MLIPs), however, can address these shortcomings due to their significant flexibility advantage over empirical potentials. Trained on quantum-accurate DFT calculations, MLIPs offer accuracy near that of first-principles methods with the computational efficiency of molecular dynamics. We can expect a MLIP to be ideal for simulating epitaxial growth, as the model must be both accurate enough to capture the complex phenomena at work, and efficient enough to simulate large-scale structures [21].

While MLIPs offer promise for modeling phenomena out of reach of traditional empirical potentials, they remain significantly more computationally expensive than empirical potentials [22], though still far more efficient than first-principles methods [23, 24]. To train an MLIP, atomic structures in the training set are first transformed into a machine-interpretable form using symmetry-invariant descriptors that encode each atom's local environment. Computing these descriptors is often costly, as it involves mathematically intensive formulations, such as those used in the Smooth Overlap of Atomic Positions representation [25]. The subsequent energy and force predictions, typically evaluated through neural networks or regressions, add further cost compared to the simple analytical forms of empirical potentials. These computational penalties make large-scale epitaxial growth simulations impractical for many current MLIPs, despite their superior accuracy in capturing the relevant atomic interactions [26, 27].

The ultra-fast force fields (UF3) machine learned interatomic potential framework bridges the gap by offering efficiency closer to that of empirical potentials while still maintaining the high accuracy associated with MLIPs [28]. UF3 accomplishes this through a cubic B-spline representation of the potential energy surface (PES) with a combination of 2-body and 3-body terms. Additionally, the environmental descriptor is efficient, as it is only based on interaction distances. UF3 has been shown to reproduce structural, mechanical, and thermal properties with accuracy compared to DFT in systems such as Tungsten or SiC [26, 28]. Additionally, UF3 accomplishes this while remaining one of the least computationally expensive MLIP frameworks [28]. The combination of speed and accuracy makes UF3 the ideal choice for our large-scale epitaxial growth potential.

This work addresses the gap in interatomic potentials for epitaxial growth of GaN on AlN by first developing a machine-learned interatomic potential for AlN using the UF3 framework. This AlN MLIP is a necessary first step toward modeling epitaxial growth of the AlN/GaN heterostructures. As we shall show, a UF3 potential trained on a diverse dataset reproduces the structural, mechanical, and thermal behaviors of AlN, including the reproduction of the wurtzite crystal structure in the overlayer of homoepitaxial growth simulations.

## II. Methodology

The development of a MLIP consists of three steps: (1) generation of reference data, (2) fitting a model to the dataset, and (3) testing and validation of the model. The following section details the methodology employed at each stage to obtain the final interatomic potential.

### A. UF3Tools Automation Framework

Due to the data-driven nature of machine-learned interatomic potentials, as well as the advanced capabilities of high-performance computing environments, the three steps of MLIP development (data generation, model fitting, and model validation) become rather complex.

These tasks are cumbersome to complete manually. Many automation workflow tools addressing one or all of the steps have been developed for other machine learned interatomic potential frameworks, such as DP-GEN, or FitSNAP [29, 30]. However, a comparable automation suite does not exist for UF3. While universal automation tools do exist, such as *wfl* [31], there are many advantages to a tightly linked ecosystem of an MLIP framework and its automation suite, such as user-friendliness and computational efficiency. We therefore developed a tightly integrated automation framework, called UF3Tools, to handle each stage of UF3 potential development, enabling rapid construction of new potentials with minimal prior expertise in DFT, machine learning, or python scripting.

The automation workflow begins by retrieving reference structures from the Materials Project Database and systematically transforming them into supercells using the Pymatgen and ASE python libraries [31, 32]. These structures are used in DFT calculations discussed in Section II.B. Each data generation step contains a variety of user-configurable settings to ensure that the data is high-quality. With this comprehensive dataset, accurate interatomic potentials can be developed that will fit a variety of use-cases. If a more complex or specific use-case is needed, such as epitaxial growth, this initial dataset can be augmented with additional DFT calculations provided by the user. The initial dataset, as well as any datasets created in the future, can easily be imported into the automation tool by simply providing the directory containing the data.

Once the dataset is complete, the next step is model fitting. The training structures are converted into UF3's cubic spline environmental descriptor through a featurization process, which encodes atomic interactions (two- and three-body) in terms of spline coefficients. This step is computationally intensive and, in traditional UF3 workflows, repeated for each trained model, even when only minor dataset modifications are applied. To avoid unnecessary precomputation, we modified the UF3 library to partition the full dataset into configurable subsets, featurize the complete dataset once, and then use the subsets for fitting. This change eliminates the most time-consuming portion of the model training workflow, as it must only be performed once, and enables rapid iteration over data filters and hyperparameters.

Further modifications were introduced to enhance parallelism. While UF3 already supported parallelized featurization, model fitting was previously constrained by single-threaded data loading and limited use of CPU cores. The revised workflow decouples ingestion of featurized data from fitting and introduces shared-memory parallelism to maximize CPU utilization. During data preparation, training data are preloaded from HDF5 files into contiguous numerical arrays using a thread pool. These arrays are then partitioned into chunks and distributed across multiple worker processes. Each worker accesses the data directly through shared memory rather than serialized copies, which avoids memory duplication and inter-process communication costs. Thread usage within each process is explicitly controlled to scale with chunk size and prevent oversubscription. Each process performs its portion of the least-squares accumulation independently, and the results are combined at the end of each batch. This approach significantly reduces fitting time and allows larger datasets to be processed more quickly by distributing workloads across additional CPU cores. These improvements are particularly beneficial for hyperparameter optimization, where many candidate models must be trained and tested.

Model validation proceeded in two stages. First, traditional error metrics (root-mean-square errors in energies and forces relative to DFT) were computed on subsets of the data. While RMSE provides an objective measure of overall fitting accuracy, we observed that, for some datasets, lower RMSE values did not necessarily translate to improved performance in simulations. Therefore, an application-testing process using LAMMPS was incorporated into the automation framework to better assess model performance. The first stage of this testing consisted of static calculations, which provided lattice constants, cohesive energies, elastic constants, and surface energies, all benchmarked against DFT. Only potentials meeting user-defined error tolerances advanced to the second step of more expensive dynamic tests. For this work, dynamic validation focused on epitaxial growth simulations; however, the automation framework supports additional types of simulations. For example, phonon properties can be evaluated using Phonopy in conjunction with UF3, and thermal expansion can be computed using MD simulations

UF3's fitting procedure contains six primary hyperparameters: three ridge regularizers (1-, 2-, and 3-body), two curvature regularizers (2- and 3-body), and a force-to-energy balance parameter. Additionally, the relative weighting of each dataset can be adjusted. For example, greater weight can be assigned to wurtzite configurations to improve near-equilibrium accuracy. Because these parameters strongly influence model accuracy, we employed the Optuna python library to efficiently search the hyperparameter space using the Tree-structured Parzen Estimator (TPE) algorithm [33], which performed the best in our testing of each of the available algorithms. To link the LAMMPS validations to the hyperparameter optimizer, a simple user-configurable loss function was defined as:

$$Loss = \sum_{i=0}^{n} w_i * \frac{error_i}{target_i} , \qquad (1)$$

where $n$ represents the total number of metrics used, $w_i$ is the user-assigned weight for the importance of metric $i$, $target$ is the target error for metric $i$, and $error_i$ is the tested percent error of this metric. This allows the optimization algorithm to prioritize metrics that may be more relevant to the user's desired application. The hyperparameter optimization in the automation tool employs a two-level parallel workflow: multiple Dask [34] workers train and test models simultaneously, and each step performed by a worker utilizes multiple cores to enhance speed. The automation tool combined with the UF3 improvements yields a robust, automated workflow for UF3 interatomic potential development. The framework enables future UF3 interatomic potentials, such as the combined Ga-Al-N epitaxial growth potential, to be rapidly developed. The tool is freely available on GitHub [35].

B. Training Data

To map the complex potential energy surface (PES) of the system, MLIPs require a diverse dataset of density functional theory (DFT) calculations. Standard MLIP training procedures rely heavily on relaxation trajectories of supercells and ab initio molecular dynamics (AIMD) simulations, which provide robust coverage of near-equilibrium states [36]. While this approach works well for bulk materials, it is poorly suited for epitaxy, where growth is inherently highly non-equilibrium, and local bonding environments can deviate substantially from those in the perfect crystal. To address this, we employed a hybrid strategy: combining conventional datasets with a large set of non-equilibrium structures generated through a genetic algorithm (GA) search [37], thereby improving representation across the full PES.

As a starting point, the known AlN polytypes, wurtzite (mp-661), rock-salt (mp-1330), and zinc-blende (mp-1700), retrieved from the Materials Project Database [38] (see Fig. 1), were expanded into 2x2x2 and 3x3x3 supercells. These supercells were used in a series of DFT calculations with the Vienna Ab Initio Simulation Package (VASP) [39, 40], using the project-augmented wave (PAW) method and the Perdew-Burke-Ernzerhof (PBE) functional [41, 42]. First, the supercells were fully relaxed to obtain the minimum-energy states. Next, systematic volume and shape perturbations were applied using VASPKIT to sample configurations near equilibrium [43]. The applied deformations included both uniform volumetric strains (±3%) and anisotropic distortions along various lattice directions, after which each structure was relaxed. These perturbed configurations capture the forces near equilibrium, ensuring the potential accurately reproduces mechanical properties. To further enrich the dataset, AIMD simulations were performed under NVT conditions at 300K, 600K, and 4000K for 5 ps, yielding perturbed local configurations relevant to finite-temperature behavior. This portion of the dataset was generated automatically using the UF3Tools automation framework developed in this work.

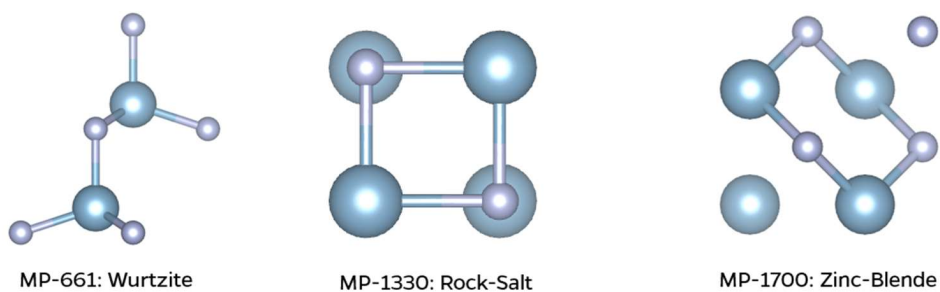

Figure 1. The three selected stable structures for AlN. Structures visualized in Vesta.[44]

Although the initial dataset provided high-quality coverage near equilibrium, it lacked representation of far-from-minimum configurations, critical for epitaxy, where adatoms encounter high-energy environments before relaxing into crystalline order. To capture these regions, we employed the Genetic Algorithm for Structure and Phase Prediction (GASP) [37], seeded with the known stable polytypes of Al, N, and AlN, to search for structures across the complete compositional space. Successive GA runs evolved increasingly diverse structures, each guided by DFT relaxations. The inclusion of both relaxed and higher-energy unrelaxed configurations broadened the PES coverage beyond standard crystal structures. Particular emphasis was placed on generating non-stoichiometric configurations, since local environments during epitaxy often deviate strongly from bulk composition. Each GA run produced thousands of candidate structures and their corresponding relaxation trajectories. However, since the focus of this work is on solid-state phases, configurations corresponding to non-solid environments, such as those containing molecular nitrogen [45], were excluded from the dataset. Additional filters were applied to eliminate unstable configurations: structures with energies per atom above 0 eV, atomic forces exceeding 25 eV/Å, or energies above the convex hull by more than 4 eV were excluded.

## III. Results & Discussion

This section will demonstrate the performance of the final UF3 MLIP in comparison to DFT and to existing interatomic potentials for AlN. Using the UF3Tools automation tool, thousands of trial potentials were trained and validated through a series of tests to compare their performance with DFT. The best interatomic potentials were used in a homoepitaxy simulation to evaluate their predictive accuracy in crystal growth.

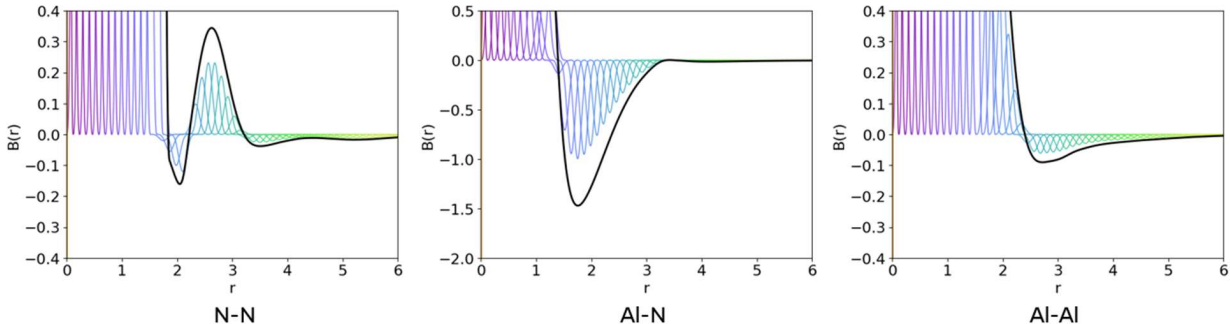

Figure 2. The two-body interaction curves from the UF3 MLIP. The overall interatomic potential is given in black. The colored lines are the spline components that sum to the total interaction potential.

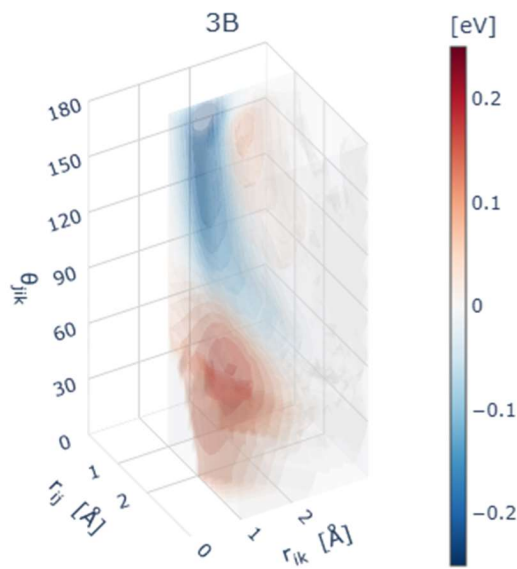

Figure 3. A plot of the three-body Al-N-N interaction from the UF3 MLIP.

A key advantage of UF3 over other MLIP frameworks is the interpretable form of its 2-body and 3-body interaction terms. This feature allows direct inspection of functional forms to diagnose deficiencies in the dataset or model training. Figure 2 illustrates the learned 2-body Al-Al, Al-N, and N-N terms. The nitrogen-nitrogen interaction is particularly notable. Because the genetic algorithm frequently generated structures containing nitrogen gas, the dataset included many short-range, strongly bonded N-N pairs. If incorporated directly, these structures biased the MLIP toward clustering nitrogen atoms, disrupting the wurtzite growth and preventing epitaxy. To resolve this, the dataset was carefully filtered by N-N pair distance, removing configurations dominated by nitrogen dimers while retaining chemically relevant configurations for crystalline AlN [45].

Figure 3 presents the learned 3-body Al-N-N interaction. The desaturated colors indicate the relatively low magnitude, on the order of 0.1 eV, relative to the dominant two-body term. This highlights the idea that the three-body contribution primarily serves as a subtle correction to stabilize the local environment, rather than to alter the overall energetic characteristics of the pairwise interactions.

1. Performance of the UF3 Model

Quantitative performance was first evaluated by comparing the UF3 predictions for per-atom energies and forces against DFT reference values. Figures 4 and 5 present these comparisons for both the GA-generated and wurtzite-specific datasets, respectively. The GA data span a wide energy range from -8 eV/atom to -3 eV/atom, with forces up to +/- 25 eV/Å, reflecting the highly diverse and often high-energy configurations encountered during structure search. The wurtzite dataset, by contrast, is narrowly distributed, with energies from -7.5 to -7.0 eV/atom and forces within +/- 1 eV/Å. The root-mean-square errors (RMSEs) reflect the distinct nature of the two datasets: 0.196 eV/atom for energy and 0.204 eV/Å for forces on the GA data, versus 0.030 eV/atom and 0.007 eV/Å for the wurtzite dataset. The errors in the GA data are consistent with other UF3 models trained on similarly diverse, high-energy configurations, while the lower errors for the wurtzite dataset align with values reported for MLIPs optimized for near-equilibrium crystalline data [15, 17, 26, 45].

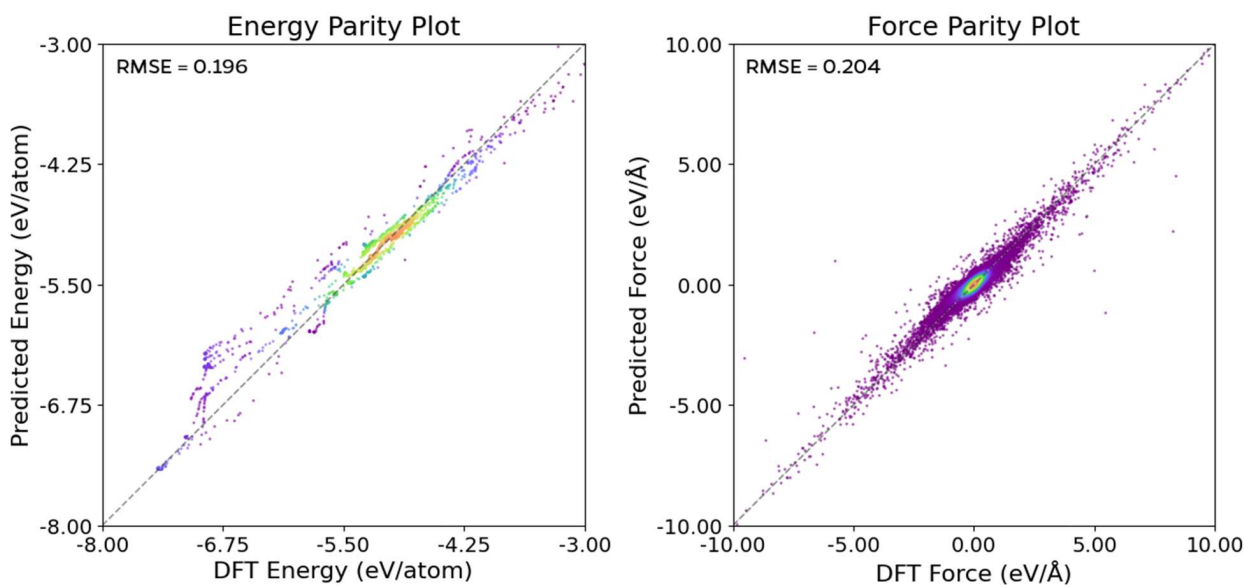

Figure 4. Comparison of UF3 predictions with reference DFT data from GA search for energy (left) and force(right).

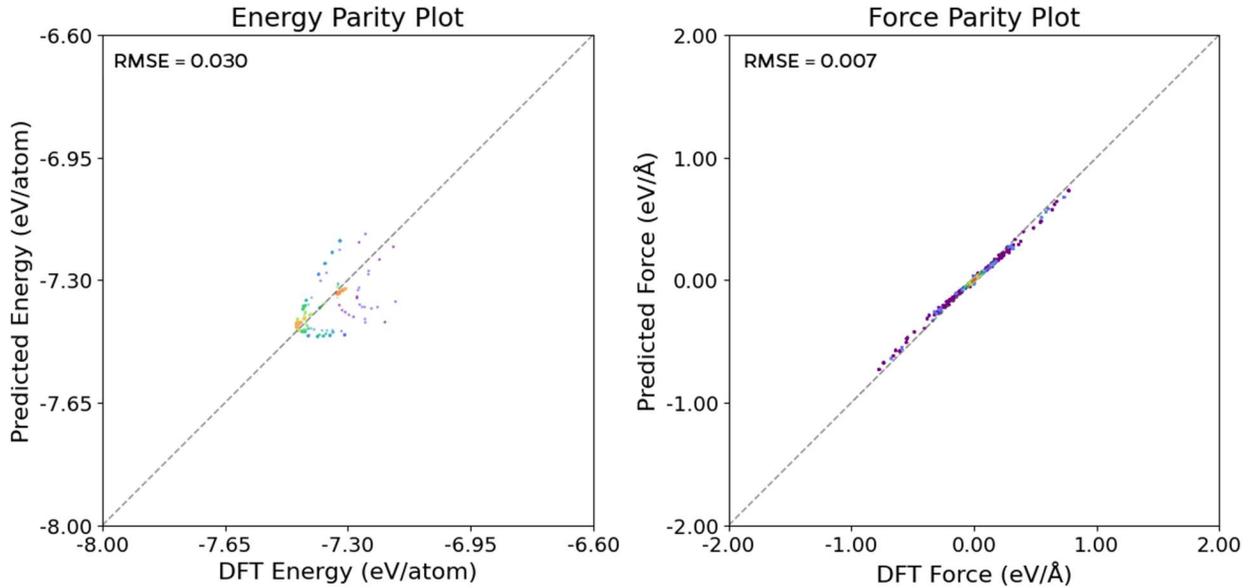

Figure 5. Comparison of UF3 predictions with reference DFT data from wurtzite data for energy (left) and force(right).

By training the UF3 model on both low-energy wurtzite structures and high-energy GA structures, the PES is comprehensively sampled across the regions relevant to homoepitaxy. This dual training strategy is essential because epitaxial growth involves atomic configurations far from equilibrium, where atoms are initially deposited in high-energy, non-crystalline environments before relaxing into the wurtzite lattice. Conventional MLIP training workflows, which rely heavily on AIMD data of crystalline structures, excel at near-equilibrium predictions but often fail to capture the PES regions critical for modeling epitaxial growth.

2. Structural and Mechanical Properties

The structural and mechanical properties predicted by each interatomic potential were benchmarked against DFT using the automated workflow described in Section II. Tables I, II, and III summarize the percent errors in lattice constants, cohesive energies, elastic constants, and surface energies across the wurtzite, rock-salt, and zinc-blende polytypes. The UF3 potential is compared against a Stillinger-Weber potential [19], two Tersoff potentials [16, 18], a Vashishta potential [46], and an Atomic Cluster Expansion potential [17].

The UF3 MLIP demonstrates strong agreement with DFT for fundamental structural and mechanical properties such as surface energies, cohesive energies, and many elastic constants, with errors typically below 10%. Elastic constants show more variation, especially in the off-diagonal terms, which are well-documented to be challenging for interatomic potentials to capture [26, 47, 48]. While the UF3 MLIP is not the most accurate at all of the properties, it achieves the best overall balance and, as we shall see, is the only potential capable of reliably capturing the crystallization relevant to epitaxial growth. Other parameterizations of the UF3 potential performed better on the metrics in Tables I-III but were not selected due to their reduced performance in epitaxy simulations. The UF3 potential, due to its highly efficient form, trades some accuracy in predicting all of these properties for stability in epitaxial growth and accurate prediction of the structure of the overlayer. Additionally, this highlights an important capability of the automation framework: a large set of plausible UF3 models can be efficiently generated and screened, allowing selection of the potential best suited to a specific target application.

Surface energies were treated with special care, since we believe they are critical to performing accurate epitaxial growth simulations. The surface energy of a polar structure, such as wurtzite [0001] AlN, is difficult to evaluate due to its asymmetric nature. In DFT calculations, a layer of atoms is typically removed from one end of the model to produce a slab structure with the same terminations on both ends. Then, the chemical potential of the missing atoms is calculated and used to produce a range of possible surface energies [49-51]. To correctly account for the depolarization field inherent to polar slabs, a dipole correction is applied within VASP. However, the use of a chemical potential is not possible in LAMMPS, where total energy is derived only from explicit atomic interactions. To enable direct comparison between MD and DFT properties, two complementary strategies were adopted.

The first strategy utilizes a stoichiometric slab. A 2x2x4 supercell of wurtzite AlN was first relaxed in its bulk configuration. Surfaces were then created by cleaving along [0001], producing a slab with one Al-terminated face and one N-terminated face. After relaxation, the surface energy was computed as

$$\gamma = \frac{E_{slab} - E_{bulk}}{2A}$$

where $E_{slab}$ is the energy of the relaxed slab, $E_{bulk}$ is the energy of the relaxed bulk supercell, and A is the area of the slab face. The UF3 potential predicted the surface energy of the stoichiometric wurtzite model with a 3.6% error relative to DFT.

The next strategy aims to compare the energies of Al- and N-terminated surfaces. One atomic layer was removed from the supercell to form a slab with symmetric terminations. Periodic boundary conditions make such structures incompatible with bulk relaxation, so the reference bulk energy was estimated from the per-atom energy of the stoichiometric bulk structure. The surface energy in this case was computed as

$$\gamma = \frac{E_{slab} - NE_{bulk}}{2A}$$

where $E_{slab}$ is the energy of the relaxed slab, $E_{bulk}$ is the energy per atom of the relaxed bulk supercell, N is the number of atoms in the slab, and A is the area of the slab face. The UF3 potential predicted surface energies for the Al- and N-terminated models with errors of 13.1% and 16.5%, respectively, in comparison to DFT.

These two strategies allow a meaningful comparison between UF3 and DFT despite the constraints of MD. The results show that the UF3 potential predicts the wurtzite [0001] surface energy with only 3.6% error, while empirical potentials deviate significantly more. This level of accuracy is critical, as surface energetics are expected to have a large influence on the dynamics of epitaxial growth.

**Table I: Percent Error of Predicted Properties for Wurtzite AlN Compared to DFT**

| Property | UF3 | Zhou SW | Karaaslan Tersoff | Tungare Tersoff | ACE | Vashishta | DFT |
|---|---|---|---|---|---|---|---|
| a=b | -3.4% | -1.6% | 8.7% | 0.4% | -0.4% | -0.7% | 3.13 Å |
| c | -2.4% | 0.3% | 10.1% | 2.3% | -0.7% | -0.7% | 5.02 Å |
| $E_{cohesive}$ | -8.6% | 1.4% | 18.6% | -3.1% | 6.4% | 0.7% | -5.77 eV |
| $C_{11}$ | 0.8% | 94.3% | -23.2% | 3.7% | -2.3% | 4.3% | 376.1 GPa |
| $C_{12}$ | -15.3% | 87.5% | -9.4% | -4.9% | -3.9% | 53.3% | 128.5 GPa |
| $C_{13}$ | 24.2% | 115.6% | -26.7% | -10.4% | 29.3% | 57.6% | 96.2 GPa |

| | | | | | | |
|---|---|---|---|---|---|---|
| $C_{33}$ | -24.5% | 108.6% | -8.6% | 16.3% | -10.3% | -32.0% | 366.3 GPa |
| $C_{44}$ | -3.1% | 99.3% | -33.4% | 5.8% | -3.2% | -18.3% | 112.3 GPa |
| Surface Energy | -3.6% | 117.8% | 74.6% | 104.9% | 47.9% | -50.2% | 237 meV/Å$^2$ |

**Table II: Percent Error of Predicted Properties for Rock-Salt AlN Compared to DFT**

| Property | UF3 | Zhou SW | Karaaslan Tersoff | Tungare Tersoff | ACE | Vashishta | DFT |
|---|---|---|---|---|---|---|---|
| a=b=c | -2.4% | 8.1% | 13.7% | 2.6% | 0.3% | 0.4% | 4.07 Å |
| $E_{cohesive}$ | -7.3% | -14.9% | 1.5% | -11.6% | 4.3% | -1.0% | -5.59 eV |
| $C_{11}$ | -18.1% | -49.8% | 28.2% | 76.7% | 41.6% | -50.7% | 424.9 GPa |
| $C_{12}$ | 27.0% | 316.9% | -100.0% | -100.0% | 166.0% | 3.0% | 166.9 GPa |
| $C_{44}$ | -13.4% | -37.6% | -86.3% | -80.1% | -14.6% | -46.6% | 306.6 GPa |

**Table III: Percent Error of Predicted Properties for Zinc-Blende AlN Compared to DFT**

| Property | UF3 | Zhou SW | Karaaslan Tersoff | Tungare Tersoff | ACE | Vashishta | DFT |
|---|---|---|---|---|---|---|---|
| a=b=c | -2.4% | -0.3% | 9.5% | 1.7% | 0.2% | 0.6% | 4.37 Å |
| $E_{cohesive}$ | -8.5% | 1.9% | 19.1% | -2.7% | 6.5% | 1.0% | -5.74 eV |
| $C_{11}$ | 10.5% | 119.0% | -26.5% | 8.2% | 1.9% | 7.6% | 299.0 GPa |
| $C_{12}$ | -20.8% | 59.2% | -22.2% | -16.6% | 3.1% | -0.5% | 164.6 GPa |
| $C_{44}$ | -0.9% | 50.3% | -28.6% | -8.3% | 9.5% | -35.6% | 185.2 GPa |

3. Dislocations

Accurate modeling of dislocation cores is essential for addressing the grand challenge of dislocation formation in epitaxy. Dislocations exhibit multiscale behavior: long-range strain fields arise from the underlying atomic-scale core, and it is this core structure that governs key dynamic phenomena, such as dislocation mobility and dissociation. These processes control strain relaxation pathways and ultimately determine the defect density in epitaxial films. If an interatomic potential fails to reproduce the correct bonding environment at the core, it can lead to spurious dislocation pathways, incorrect formation energies, or incorrect predictions of dislocation behavior during growth. A reliable potential must therefore capture both bulk elastic properties and the detailed atomistic structure of the core.

In wurtzite AlN, dislocations typically have Burgers Vectors of $\frac{1}{3}\langle 11\bar{2}0\rangle$ for edge type, $\langle 0001\rangle$ for screw type, or $\frac{1}{3}\langle 11\bar{2}3\rangle$ for mixed edge and screw type [52, 53]. These dislocations preferentially align along the [0001] lattice direction. Experimental studies of group-III nitrides consistently identify two equilibrium core structures: the 8-atom ring and the higher-energy 5/7-atom ring [54]. To test the UF3 potential's fidelity, a simulation cell of wurtzite AlN was constructed. A dipole of dislocations with Burgers vector $\frac{1}{3}\langle 1\bar{2}10\rangle$ was inserted along the [0001] direction, separated by 40 unit cells. The dipole was necessary to satisfy periodic boundary conditions. The structure was relaxed through energy minimization to obtain the lowest-energy configuration of the dislocation cores. The relaxed core structure is shown in Fig. 6.

The UF3 potential predicts the experimentally observed 8-atom ring core, in contrast to many interatomic potentials that stabilize a 4-atom ring configuration [55]. The latter has appeared frequently in MD simulations but is rarely seen in experimental studies. Earlier iterations of the UF3 potential also exhibited this 4-atom core, but this issue was resolved through more extensive hyperparameter optimization. The final parameterization achieved higher accuracy and better transferability than the unoptimized model, highlighting the important role of hyperparameters in MLIPs. The agreement with experimental dislocation core structures, combined with UF3's accurate elastic constants, suggests that the potential can reliably describe dislocation behavior during epitaxial growth.

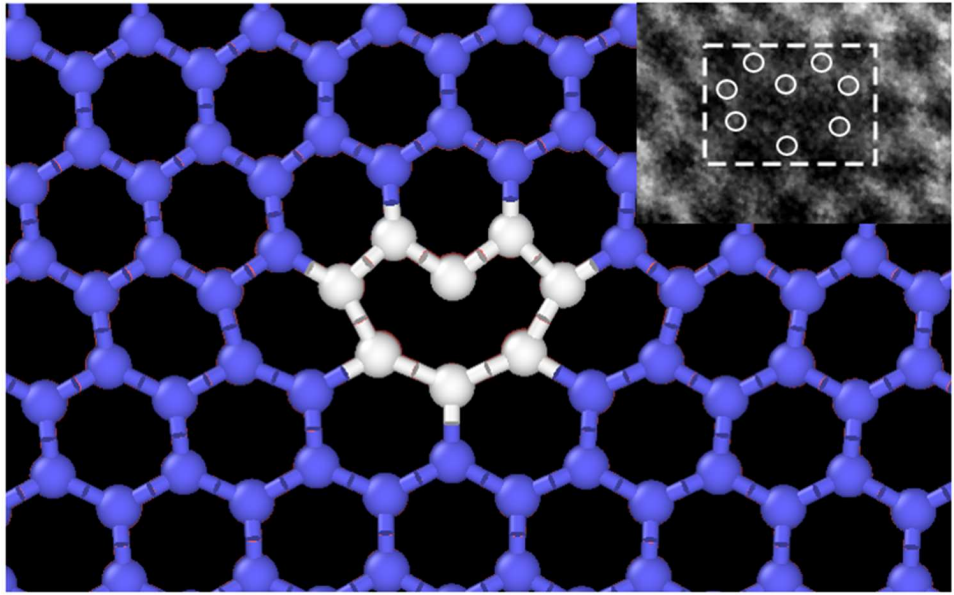

Figure 6. The predicted dislocation core structure, visualized in OVITO. Experimental result is shown in the top right for comparison [54]. White circles have been overlaid on the experimental image to indicate the atomic positions.

In addition to the dislocation core structure, it is important that the potential can predict the energy associated with certain point defects. To evaluate this, supercells consisting of 3 by 3 by 3 wurtzite unit cells were constructed, each containing a distinct defect configuration. A single Frenkel, Schottky, or antisite defect was introduced into the model using the Pymatgen library. The defect formation energies were computed by first relaxing the defect-free bulk supercell using both DFT and LAMMPS. Each defective configuration was then relaxed, and the defect energy was determined as the difference in energy per atom between the defective and pristine supercells. As with the surface energies, the chemical potentials of the missing atoms cannot be accounted for within LAMMPS, as energy is defined only by atomic interactions. The energies associated with each defect were obtained from the energy difference between the defective and pristine structures. A parity plot comparing the defect formation energies predicted by the UF3 potential against the corresponding DFT results is presented in Fig. 7.

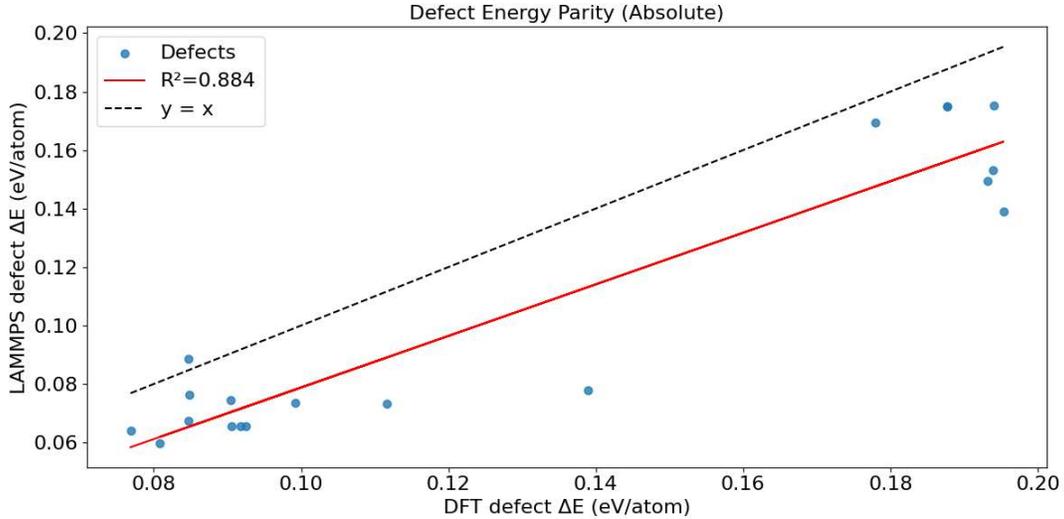

Figure 7. A comparison between DFT and the UF3 potential of the change in energy per atom associated with a defect.

These results demonstrate that the UF3 potential reproduces a broad range of AlN properties with high fidelity, from bulk elastic behavior to atomically resolved defect structures. The potential can capture both the equilibrium and non-equilibrium properties, such as predicting the dislocation core structure and providing defect energetics in reasonable agreement with DFT. Importantly, this indicates that the UF3 potential not only fits static DFT benchmarks but also preserves physically meaningful bonding environments across diverse configurations. These validations establish a strong foundation for dynamic simulations of epitaxial growth, where the potential must accurately describe far-from-equilibrium processes such as adatom diffusion, island coalescence, and defect nucleation.

4. Epitaxial Growth

The simulation of epitaxial growth in group-III nitride systems is essential for addressing the two grand challenges of defect formation and self-heating facing the development of electronics. A critical benchmark for these interatomic potentials is whether they can reproduce experimentally observed crystallization of the epilayer. None of the existing interatomic potentials that we have tested have achieved this: most produced a mixture of wurtzite and zinc-blende phases during growth, while others exhibited unphysical

To evaluate the UF3 potential, a wurtzite AlN substrate containing 189,000 atoms was constructed with dimensions of 25x25x4 nm, with its [0001] surface exposed, consistent with experimental configurations. The substrate was first relaxed by energy minimization to determine the lattice constant at 0K. It was then equilibrated under the NPT ensemble at 2000 K and 0 bar. While this temperature is slightly higher than typical MOCVD or MBE growth conditions [56], it is necessary in MD to accelerate surface diffusion and compensate for the artificially high deposition rates imposed by nanoscale simulation [57]. The resulting thermally expanded lattice constant was used to construct a new substrate, ensuring that the epitaxial simulations were performed with the accurate lattice parameters at the growth temperature.

During deposition, the bottom layer of the substrate was fixed to prevent translation of the system. Periodic boundary conditions were applied along the X ([11$\bar{2}$0]) and Y ([1$\bar{1}$00]) directions, while a free surface was maintained along the Z ([0001]) direction. To control substrate temperature, a 10 Angstrom-thick Langevin thermostat region was defined at the substrate center, held at 2000K, while the remainder of the system evolved under Newtonian dynamics. A timestep of 0.001 ps was used. Growth was modeled by randomly generating equal numbers of Al and N atoms above the substrate surface, with each atom assigned an initial velocity drawn from a Gaussian distribution centered at the growth temperature. Velocities were distributed among all spatial directions to introduce variation in the deposition angle. Each deposition cycle consisted of 20 ps of deposition followed by 30 ps of equilibration under the NVE ensemble, allowing the deposited atoms to diffuse and incorporate. The overall growth rate was approximately 0.3 monolayers per nanosecond. Although this is much higher than experimental rates, this accelerated kinetics is unavoidable at the scales of atomistic simulations [57]. Growth continued until an epilayer thickness of 3 ML was achieved, as shown in Fig. 7.

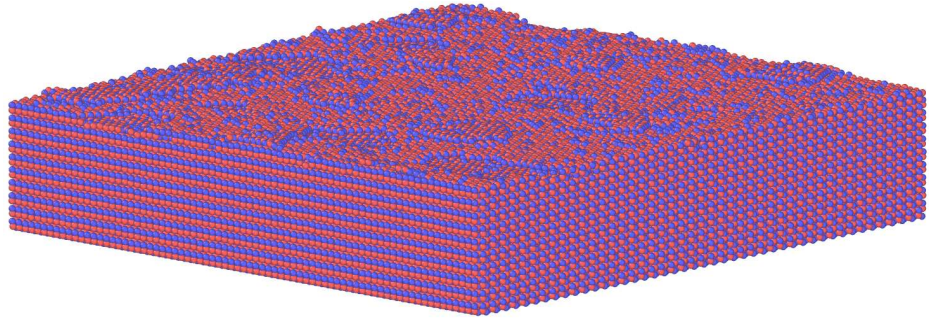

Figure 7. The grown epitaxial structure, visualized in OVITO [58]. This structure includes both the substrate and overlayer.

As can be seen from Fig.7, the UF3 simulation produced a smooth epilayer with large terraces separated by one- to three-unit cell steps, characteristic of a Frank-van der Merwe two-dimensional growth mode. A height map of the surface is shown in Fig. 8 alongside experimental images, demonstrating qualitative agreement in growth morphology [56]. Although the simulated system spans a much smaller spatial domain than experiments, the smooth, terraced epitaxial surface appears similar to the morphological trends observed experimentally.

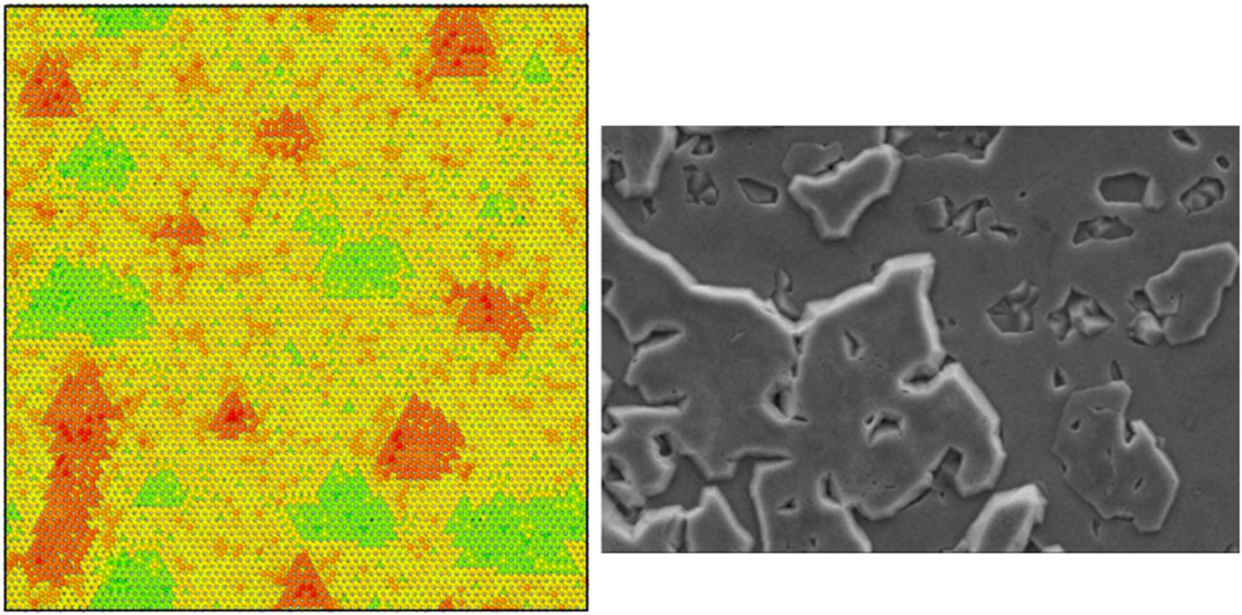

Figure 8. A topographical map of the surface of the grown overlayer. Experimental result is shown for comparison [56].

Structural analysis of the overlayer was performed using OVITO's Identify Diamond Structure algorithm, as shown in Fig. 9 [59]. The epilayer growth with UF3 crystallized into the wurtzite phase, consistent with experimental observations. Small deviations were found at the free surface, where incomplete bonding naturally prevents full wurtzite coordination. The overlayer maintained near-100% wurtzite structure aligned with the substrate.

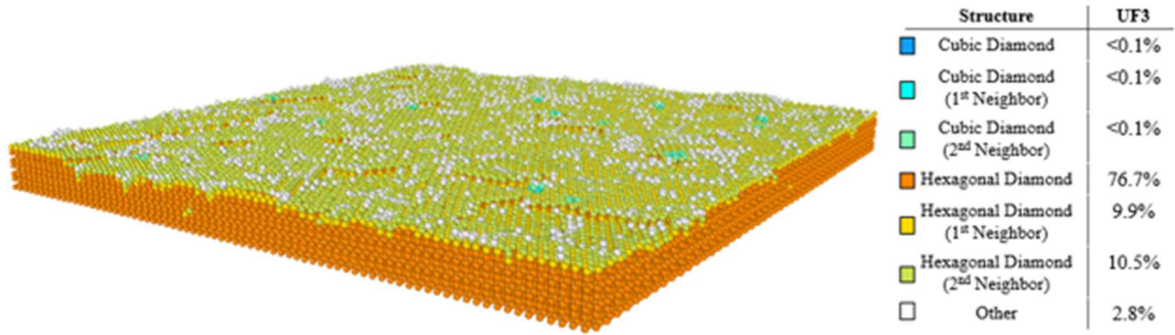

Figure 9. The overlayer obtained in homoepitaxial growth of AlN on AlN. The orange atoms indicate wurtzite structure, the blue atoms indicate zinc-blende, and the white atoms indicate an amorphous structure, according to OVITO's Identify Diamond Structure algorithm [59].

In comparison, the best-performing prior potential from the literature, the Stillinger-Weber (SW) GaN/AlN model developed by Zhou et al, failed to reproduce the correct crystallization during epitaxy. A similar epitaxial growth simulation using the Zhou SW potential, shown in Fig. 10, produced an epilayer with many large voids, and the overlayer crystallized into a mixed phase of ~60% zinc-blende and ~40% wurtzite. Such a result is consistent with prior reports [20], even after extensive testing of kinetic parameters. In these simulations, the zinc-blende structures occurred with sharp transitions to wurtzite, creating internal phase boundaries not observed experimentally.

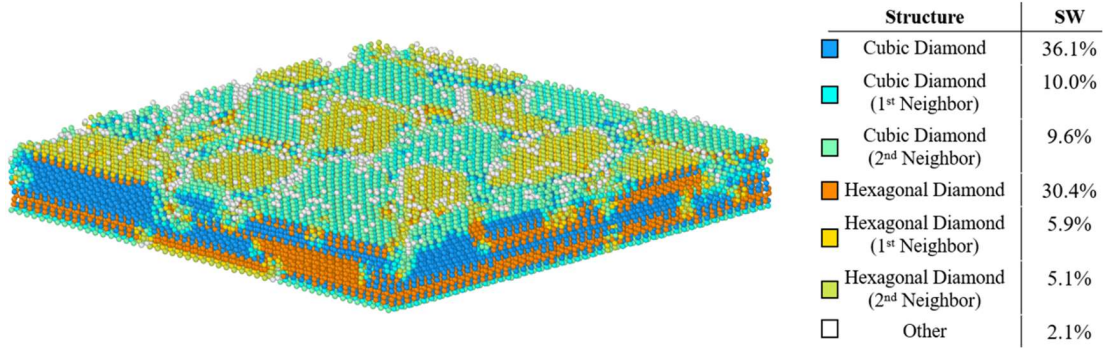

Figure 10. The overlayer obtained in homoepitaxial growth of AlN using the SW potential by Zhou et al. The structure is approximately 60% zinc-blende and 40% wurtzite, according to OVITO's Identify Diamond Structure algorithm [59].

The systematic failure of existing potentials to stabilize wurtzite in the overlayer may originate from their inability to capture the small energy difference between wurtzite and zinc-blende AlN. DFT calculations predict this difference to be approximately 0.02 eV/atom in favor of wurtzite, yet many empirical models either over-stabilize the zinc-blende phase or treat the two structures as energetically equivalent. In contrast, the UF3 potential reproduces the correct energetic ordering, predicting the wurtzite phase to be 0.016 eV/atom more stable than zinc-blende. This difference enables UF3 to stabilize the wurtzite structure during growth, predicting epitaxial films consistent with experimental observations.

## IV.     Summary and Discussions

In this work, a machine learned interatomic potential for AlN was developed for large-scale predictive simulations using simulation tools such as MD and CAC [60-62]. The potential demonstrates strong agreement with DFT in predicting key structural and mechanical properties, including lattice constants, cohesive energy, and surface energy across the wurtzite, rock-salt, and zinc-blende polytypes. It also accurately reproduces the experimentally observed atomic core structure of edge dislocations, which is critical for modeling strain relaxation through misfit and threading dislocations in epitaxial films. Most significantly, it reproduces the experimentally observed wurtzite crystal structure in the overlayer during homoepitaxial growth of AlN on wurtzite AlN, something that prior potentials failed to achieve. Additionally, the potential reproduces the experimentally observed 2D layer by layer growth mode in AlN epitaxy.

Despite its overall accuracy, the potential exhibits several notable limitations. It underperforms in predicting the elastic constants $C_{13}$ and $C_{33}$, an issue that has been documented in the literature for both machine-learned and empirical potentials [26, 47, 48]. The model also struggles with the formation energies of certain point defects. These deficiencies likely stem from the finite complexity of the UF3 functional form. While the UF3 framework offers exceptional computational efficiency for a machine-learned potential, over 5 times faster than the P-ACE potential for AlN, the limited number of spline knots restricts its ability to perfectly resolve all regions of the potential energy surface. During development, several models that performed exceptionally on benchmarks later failed in epitaxial growth simulations, underscoring that optimizing one region of the PES can degrade accuracy elsewhere due to the interdependence of the cubic B-spline representation. Because linear regression potentials can only minimize but not eliminate global error, priority in development was ultimately placed on reproducing epitaxial growth behavior.

Despite these limitations, the developed potential is expected to predict critical mechanical and microstructural processes in AlN, including fracture and dislocation dynamics, as well as their coupling, enabling predictive simulation of defect formation and evolution in AlN epitaxial processes. The combination of accuracy, transferability, and computational speed afforded by the UF3 framework makes large-scale, atomically informed simulations of epitaxial growth of AlN feasible. This development thus represents a crucial first step towards fully predictive simulations of the heteroepitaxy of GaN/AlN heterostructures, a quantitative and mechanistic understanding of the mechanisms underlying defect formation in GaN/AlN epitaxy, and co-design of Gan/AlN heterostructure and manufacturing processes.

## Acknowledgement

This work is based on research supported by the US National Science Foundation under Award Number CMMI- 2349160. The work of Chen and Phillpot is also partially supported by DMR 2121895. The computer simulations are funded by the Advanced Cyberinfrastructure Coordination Ecosystem: Services & Support (ACCESS) allocation TG-DMR190008

## Data Availability

Data will be made available on request. The UF3Tools code can be found at: https://github.com/uf-chenlab/